\documentclass[conference]{IEEEtran}
\usepackage[pdftex]{graphicx}
\usepackage{amsmath} 
\usepackage{amssymb} 
\usepackage{array}
\usepackage{blindtext}
\usepackage{amssymb}
\usepackage[utf8]{inputenc}
\usepackage[T1]{fontenc}
\usepackage{balance} 
\usepackage[numbers]{natbib}
\usepackage{bigstrut}
\ifCLASSOPTIONcompsoc
\usepackage[caption=false, font=normalsize, labelfont=sf, textfont=sf]{subfig}
\else
\usepackage[caption=false, font=footnotesize]{subfig}
\fi
\def\BibTeX{{\rm B\kern-.05em{\sc i\kern-.025em b}\kern-.08em
    T\kern-.1667em\lower.7ex\hbox{E}\kern-.125emX}}
\makeatletter
\def\ps@IEEEtitlepagestyle{%
	\def\@oddfoot{\mycopyrightnotice}%
	\def\@evenfoot{}%
}
\def\mycopyrightnotice{%
	{\hfill \footnotesize 978-1-7281-9365-6/20/\$31.00 \copyright 2020 IEEE\hfill}
}
\makeatother
\begin{document}
\title{Design and implementation in USRP of a preamble-based synchronizer for OFDM systems}

\author{\IEEEauthorblockN{Jos\'e Herrera-Bustamante,
Vanessa Rodr\'iguez-Ludeña, A.G. Correa-Mena and
Diego Barrag\'an-Guerrero}
\IEEEauthorblockA{Departamento de Ciencias de la Computación y Electrónica (DCCE),
Universidad T\'ecnica Particular de Loja-Ecuador\\
Email: jeherrera6@utpl.edu.ec,
vsrodriguez@utpl.edu.ec,
agcorreax@utpl.edu.ec,
dobarragan@utpl.edu.ec}}

\maketitle

\begin{abstract}
The Orthogonal Frequency Division Multiplexing (OFDM) is one of the most widely adopted schemes in wireless technologies such as Wi-Fi and LTE due to its high transmission rates, and the robustness against Intersymbol Interference (ISI). However, OFDM is highly sensitive to synchronism errors, which affects the orthogonality of the carriers. We analyzed several synchronization algorithms based on the correlation of the preamble symbols through the implementation in Software-Defined Radio (SDR) using the Universal Software Radio Peripheral (USRP). Such an implementation was performed in three stages: frame detection, comparing the autocorrelation output and the average power of the received signal; time synchronism, where the cross-correlation based on the short and long preamble symbols was implemented; and the frequency synchronism, where the Carrier Frequency Offset (CFO) added by the channel was detected and corrected. The synchronizer performance was verified through the USRP implementation. The results serve as a practical guide to selecting the optimal synchronism scheme and show the versatility of the USRP to implement digital communication systems efficiently.
\end{abstract}

\renewcommand\IEEEkeywordsname{Keywords}

\section{Introduction}
The Orthogonal Frequency Division Multiplexing (OFDM) technique is one of the most common modulation schemes characterized by dividing the available wireless channel bandwidth into several orthogonal subcarriers \cite{Boter_2009}. OFDM also provides low Intersymbol Interference (ISI) \cite{ElHajjar_2007}, efficient use of the spectrum, and a simpler equalization. Thus, OFDM guarantees robustness communication in both Additive White Gaussian Noise (AWGN) channels and multi-path Rayleigh channels. Due to these characteristics, OFDM has been adopted in the physical layer of the IEEE 802.11 standard and its amendments. However, a precise synchronism must be achieved to maintain the mentioned characteristics. 

Synchronism is a crucial step to receive transmitted information correctly. In an OFDM system, it is divided into three stages: frame detection, time synchronism, and frequency synchronization. The frame detection is carried out by comparing the autocorrelation of the received signal against its average power; the time synchronism determines the beginning of the information in the frame through cross-correlation of the received signal with a training sequence (i.e., short and long preamble symbols); and, finally, the frequency synchronism is used to align the frequency between the carriers of the receiver (RX) and the transmitter (TX), reducing the Inter-Carrier Interference (ICI). The last stage detects the phase of the received signal through a correlation-based scheme, where that phase is used to correct the Carrier Frequency Offset (CFO) introduced by the wireless channel \cite{Wensheng2009}.

In that regard, we implemented an effective synchronism solution for multi-carrier communication systems in a Software Defined Radio (SDR) platform, i.e., the Universal Software Radio Peripheral (USRP) \cite{Reis_2012}. In particular, we investigated the performance of the preamble-aided synchronization algorithms in an SDR environment, using GNU Radio. Together with the Ettus USRP N210 equipment capabilities, GNU Radio becomes a robust platform to simulate, test, and improve communication systems' critical stages.

This paper is organized as follows. In Section \ref{fig:setting_the_environment} are summarized the main characteristics of the OFDM technique, as well as, the particularities of the IEEE 802.11a preamble and the channel model. In sections \ref{fig:frame_detection}, \ref{fig:time_synchronism}, and \ref{fig:CFO_detector} the three stages to perform the preamble-based synchronizer design are described: frame detection, time synchronism, and frequency synchronism, respectively. In Section \ref{fig:results_and_discussion}, the results and discussion of the performance of the synchronization algorithms, in terms of the variance, are shown. Finally, in Section \ref{section:sec_conclu} the conclusions are presented. 

\section{Setting the environment}
\label{fig:setting_the_environment}
\subsection{Preamble-aided synchronization: the IEEE 802.11 standard}
In a multi-carrier communication system, the synchronization stage is usually carried out through the correlation of the preamble symbols. Such a scheme is known as preamble-aided synchronization. Some IEEE 802.11 standards have adopted a deterministic preamble to facilitate packet synchronization. Throughout this paper, we use the well-known IEEE 802.11a preamble sequence to simulate and implement the algorithms in the USRP. 

The IEEE 802.11a standard and its amends are used in Wireless Local Areas Networks (WLANs), which support speeds up to 54 Mbps, and employ OFDM as the modulation technique due to its optimum performance in highly dispersive channels. Moreover, it uses 52 subcarriers out of 64 available to deal with the adjacent channel interference (ACI), 48 of them convey user data, 4 are pilot tones for phase tracking, and the remaining 12 are null tones \cite{LAN/MANStandardsCommittee_1999}. 

\subsection{Preamble and channel model}
\begin{figure}[!b]
  \centering
  \includegraphics[width=\columnwidth]{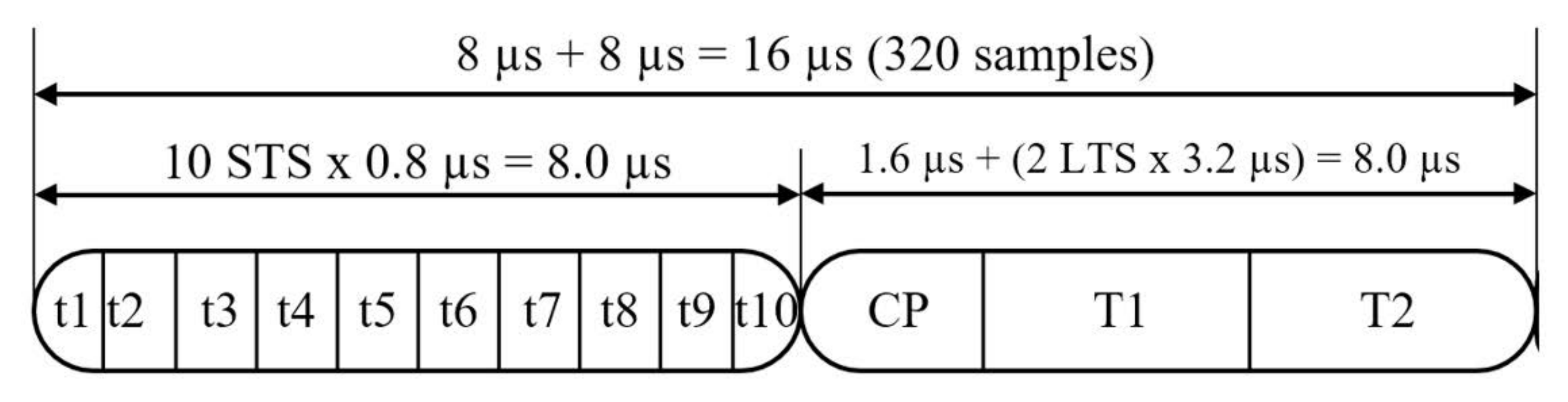}
  \caption{OFDM preamble sequence compose of STS and LTS symbols, used before data frame.}
  \label{fig:preamble}
\end{figure}

As shown in Fig. \ref{fig:preamble}, the OFDM preamble sequence is composed of two sets of short and long symbols. The ten short symbols, consisting of 16 complex samples, are known as the Short Training Sequence (STS). The two long symbols, composed of 64 complex samples, represent the Long Training Sequence (LTS) \cite{Zhou_2004}. Usually, the STS is used in frame detection and the LTS in time and frequency synchronism. Between the STS and LTS are added 32 Cyclic Prefix (CP) samples, reaching 320 samples in the full-frame. The OFDM scheme, aided with this preamble, reduces the ISI effect and makes the system robust against the multipath fading.

\begin{figure}[!t]
  \centering
  \includegraphics[width=\columnwidth]{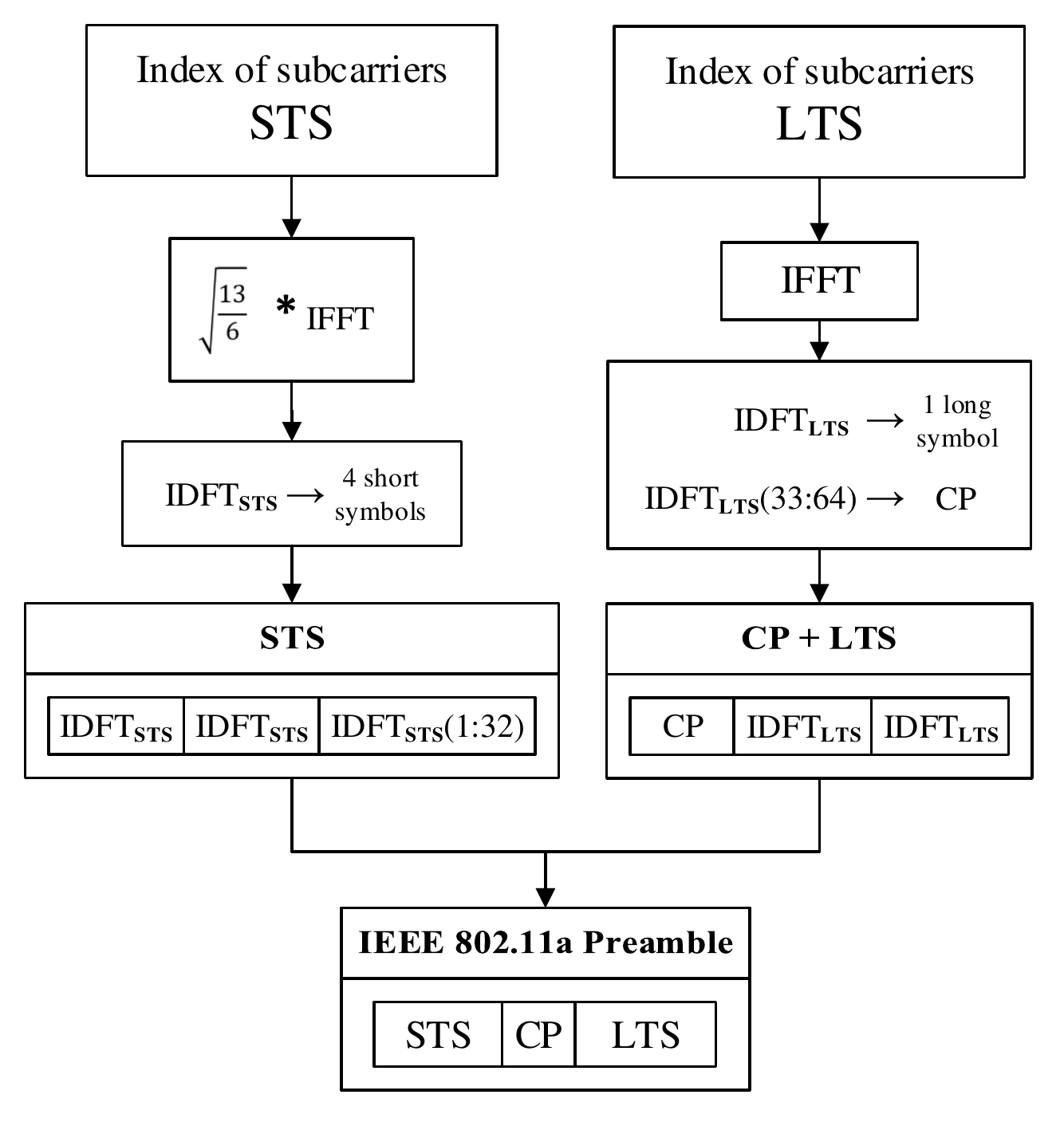}\vspace*{-1.0em}
  \caption{IEEE 802.11a preamble generation scheme.}
  \label{fig:preamble_generation}
\end{figure}

In Fig. \ref{fig:preamble_generation} the procedure to generate the IEEE 802.11a preamble is shown. The preamble construction method detailed in \cite{LAN/MANStandardsCommittee_1999} is used to generate a sequence in Python.

\begin{figure}[!t]
  \centering
  \includegraphics[width=0.8\columnwidth]{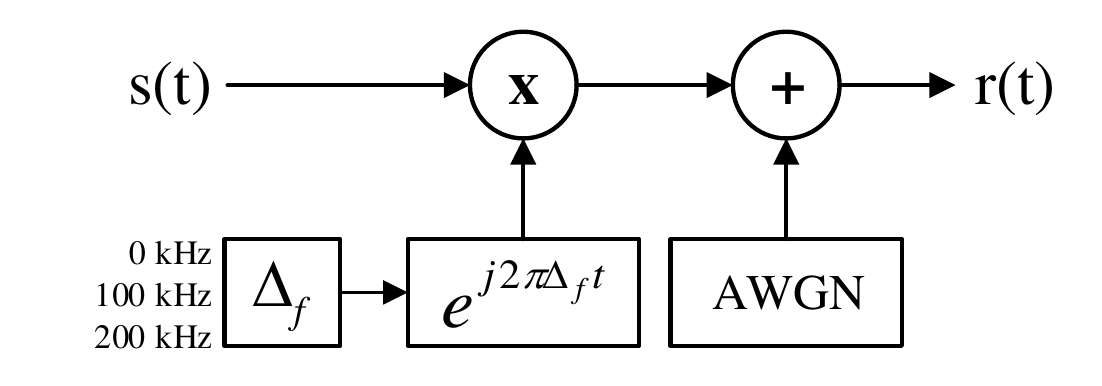}\vspace*{-0.5em}
  \caption{AWGN channel model including CFO.}
  \label{fig:channel_model}
\end{figure}

The channel model scheme is shown in Fig. \ref{fig:channel_model}, where $s(t)$ represents the transmitted signal, and $r(t)$ is the received signal affected by the channel noise and the frequency offset ${\Delta _f}$. The so-called tapped-delay channel has been selected as the channel model for this work, in particular, the A and C versions of the European Telecommunications Standards Institute (ETSI) channel models \cite{Medbo_1998}. These models include the multipath fading behavior, frequency deviation, and AWGN channel. For simulation purposes, we added the frequency offset by multiplying the transmitted signal by a complex exponential, where the value of each offset represents the optimal (${\Delta _f} = 0$  kHz), moderate (${\Delta _f} = 100$ kHz) and severe (${\Delta _f} = 200$ kHz) conditions.

\subsection{GNU Radio blocks and system setup}
All synchronism stages were designed using Python and embedded into a custom block denominated Out-Of-Tree (OOT) module inside the GNU Radio Companion (GRC) interface. Each OOT block, created through the \textit{gr\_modtool} performs dedicated processing on the received preamble. The \textit{Vector Source} provides the values obtained in the preamble generation to be transmitted and the \textit{WX GUI Scope Sink} graphically shows the response over time of each detector.

\begin{figure}[b]
  \centering
  \includegraphics[width=\columnwidth]{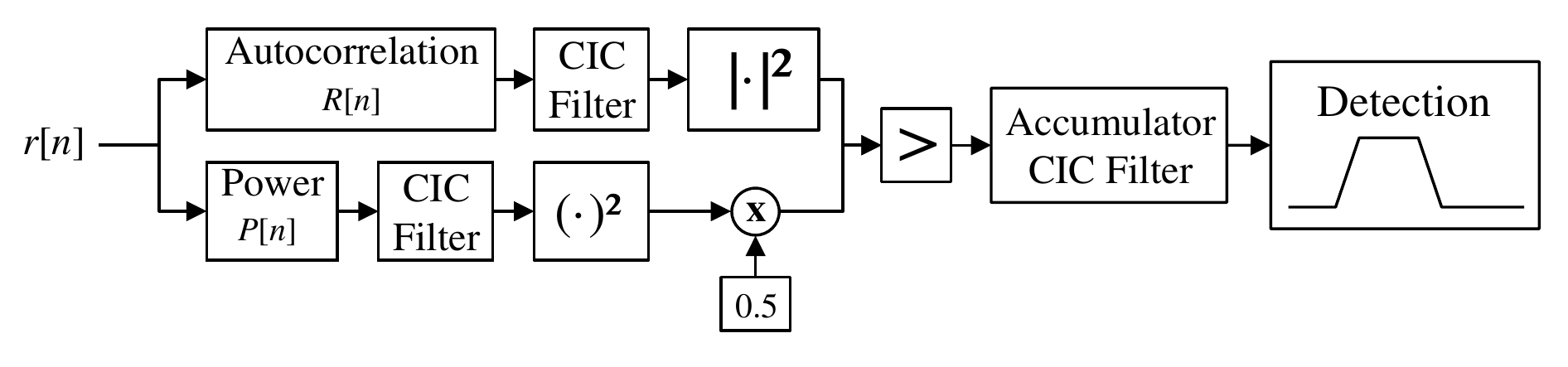}
  \caption{Frame Detection algorithm scheme.}
  \label{fig:frame_det_scheme}
\end{figure}

\begin{figure}[t] 
  \centering
  \subfloat[\label{fig:time_synch_sch_STS}]{
  \includegraphics[width=0.8\columnwidth]{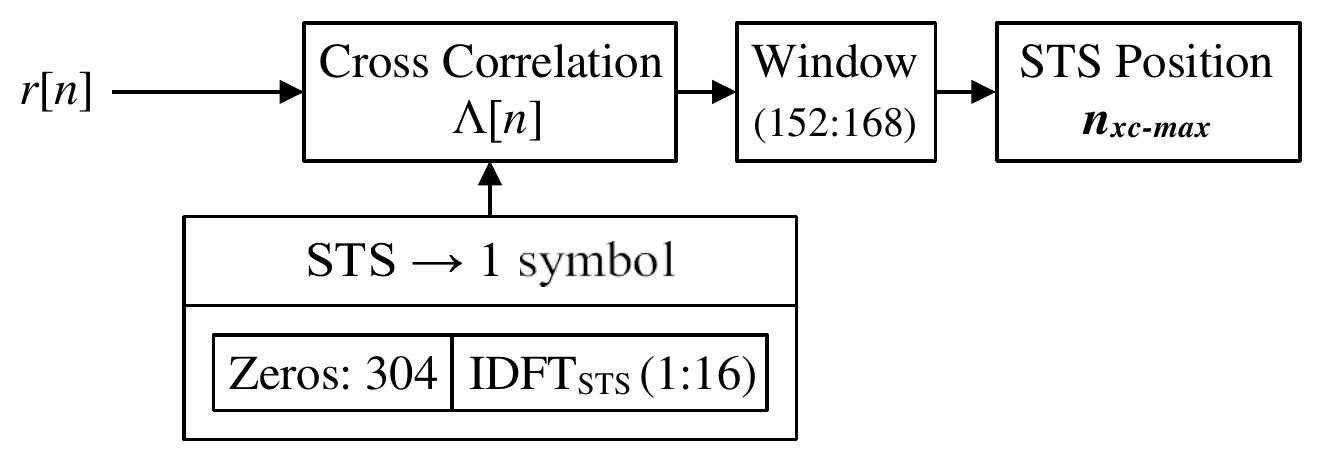}}\hfill
  \subfloat[\label{fig:time_synch_sch_LTS}]{
  \includegraphics[width=0.8\columnwidth]{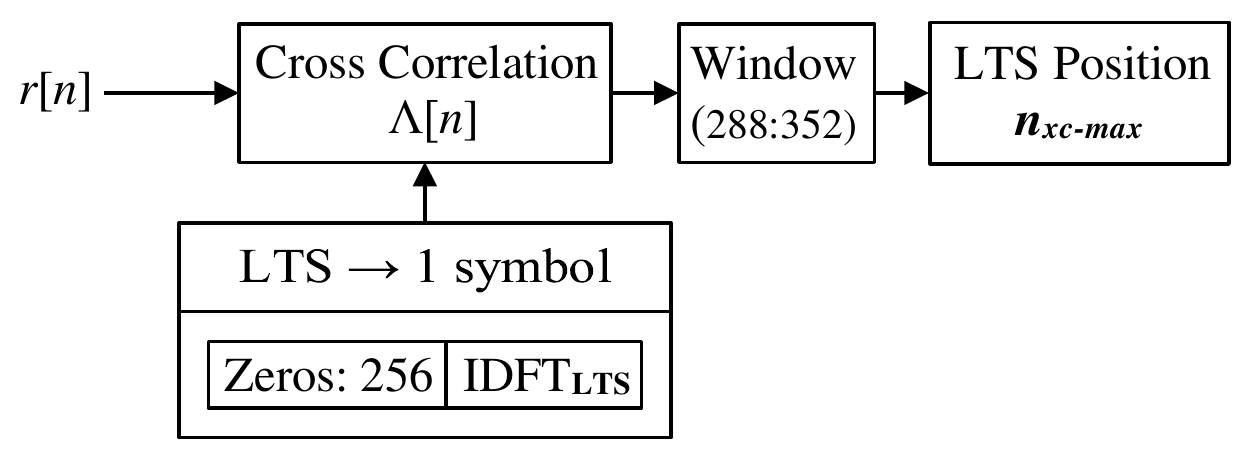}}\hfill
  \caption{Time Synchronization algorithm schemes for: (a) STS and (b) LTS.}
  \label{fig:time_synch_sch} 
\end{figure}

\begin{figure}[t]
  \centering
  \includegraphics[width=\columnwidth]{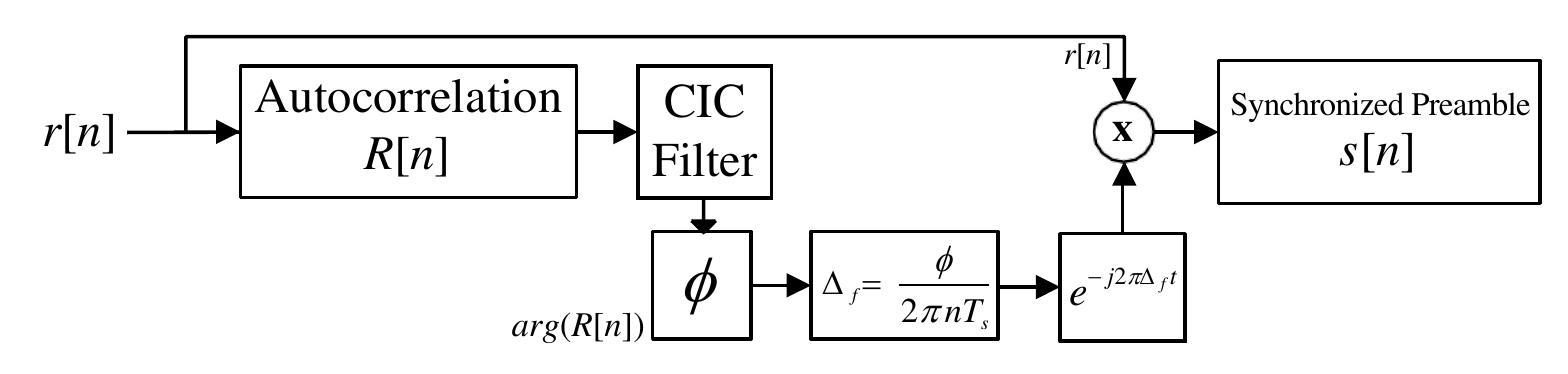}
  \caption{CFO Detection algorithm scheme with preamble synchronization.}
  \label{fig:CFO_scheme}
\end{figure}

Frame Detector detects the presence of the package. Its implementation is based on Fig. \ref{fig:frame_det_scheme}.
Time synchronism based on STS and LTS symbols is implemented according to schemes shown in Fig. \ref{fig:time_synch_sch_STS} and Fig. \ref{fig:time_synch_sch_LTS}, respectively.
CFO Detector estimates and compensates the magnitude of CFO introduced by the channel. Such detection is based on Fig. \ref{fig:CFO_scheme}.
In all the schemes, the Cascaded Integrator–Comb (CIC) works as a moving average filter. In this paper, the implemented algorithms are based on the correlation of the preamble symbols. For test purposes, only LTS and STS were sent between TX and RX; thus, the preamble is transmitted repeatedly from one USRP to another, running the implemented algorithms in each transmission. 

All synchronization algorithms were implemented in a transceiver system mounted on two PCs with Ubuntu 14.04 operating system, and its Gigabit Ethernet ports were connected to USRP N210 modules. WSS016 omnidirectional antennas were used for the radio interface. A 900 MHz frequency carrier was set up, a transmitter gain of 30 dB, a sampling frequency of 20 MHz, and BPSK signaling as the digital modulation scheme. 

\section{Frame detection}
\label{fig:frame_detection}
Frame detection informs the presence or absence of a frame (user information) at the receiver. The frame detection method implemented in this work is based on the Schmidl and Cox algorithm \cite{Schmidl_1997}, where the received signal $r_n$ is correlated with its delayed conjugate version $r_n^*$. The delay length (in samples) is equal to a short symbol duration. Thus, the correlation is denoted by:

\begin{equation}
R[n] = \sum\limits_{m = 0}^{L - 1} {\left( {{r_{n + m}} r_{n + m + L}^*} \right),} 
\label{eq:r_n}
\end{equation}
where $L$ represents the delay (16 samples, corresponding to a short symbol length).

A frame is detected if the ratio $M[n]$ between the square of the absolute value of $R[n]$ and the signal power $P[n]$ reaches a threshold. These parameters are defined as follows:

\begin{equation}
P[n] = \sum\limits_{m = 0}^{L - 1} {{{\left| {{r_{n + m + L}}} \right|}^2},} 
\label{eq:p_n}
\end{equation}

\begin{equation}
M[n] = \frac{{{{\left| {R[n]} \right|}^2}}}{{P{{[n]}^2}}}> \text{threshold.}
\label{eq:m_n}
\end{equation}

According to \eqref{eq:r_n}, the correlation output $R[n]$ is a complex parameter. To compute its modulus we have applied an approximation where ${\left| {R\left[n \right]} \right|^2}$ is calculated by: 

\begin{equation}
{\left| {R[n]} \right|^2} \approx \left| {{\mathop{\rm \mathcal{R}}\nolimits} \left\{ {R[n]} \right\}} \right| + \left| {{\mathop{\rm \mathcal{I}}\nolimits} \left\{ {R[n]} \right\}} \right|,
\label{eq:rn_simpl}
\end{equation}
where $\mathcal{R}\left\lbrace \cdot \right\rbrace$ and $\mathcal{I}\left\lbrace \cdot \right\rbrace$ denote the real and imaginary part of a complex number, respectively. In \cite{Canet_2004}\cite{Yip_2002}, it has shown that an optimal threshold for frame detection is equal to the power of the received signal divided by two. Rewriting equation \eqref{eq:m_n}, we have:

\begin{equation}
\left| {R{{[n]}}} \right|^2 > 0.5 P[n]^2.
\label{eq:rn_thr}
\end{equation}

Such a simplification has an additional advantage because a division by two is straightforwardly implemented as a shift register, optimizing the hardware resources and reducing processing latency.

\begin{figure}[b] 
  \centering
  \subfloat[\label{fig:frame_det_output_A}]{
  \includegraphics[width=\columnwidth]{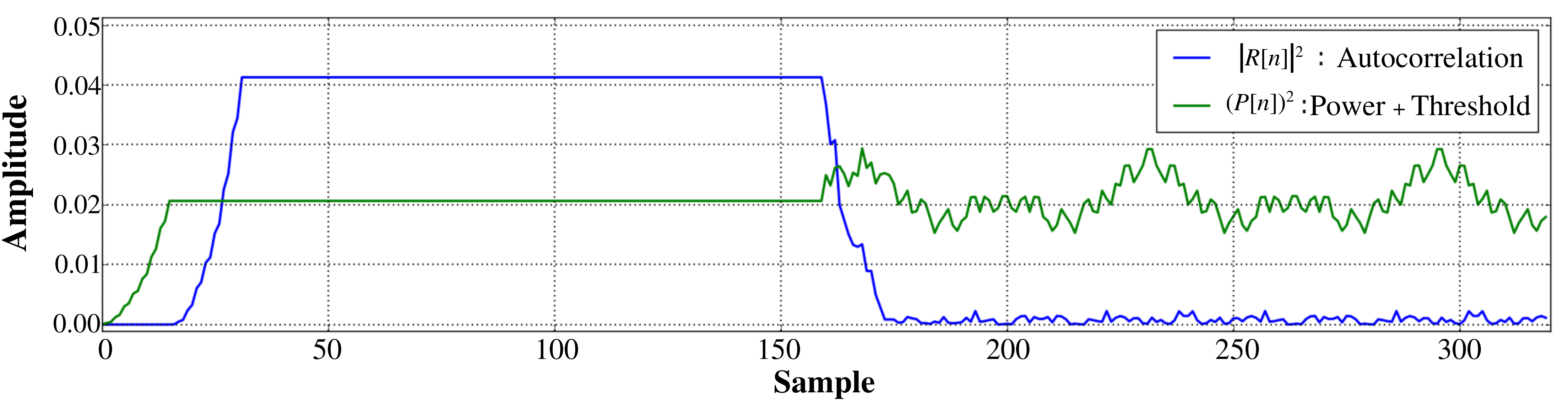}}\hfill
  \subfloat[\label{fig:frame_det_output_B}]{
  \includegraphics[width=\columnwidth]{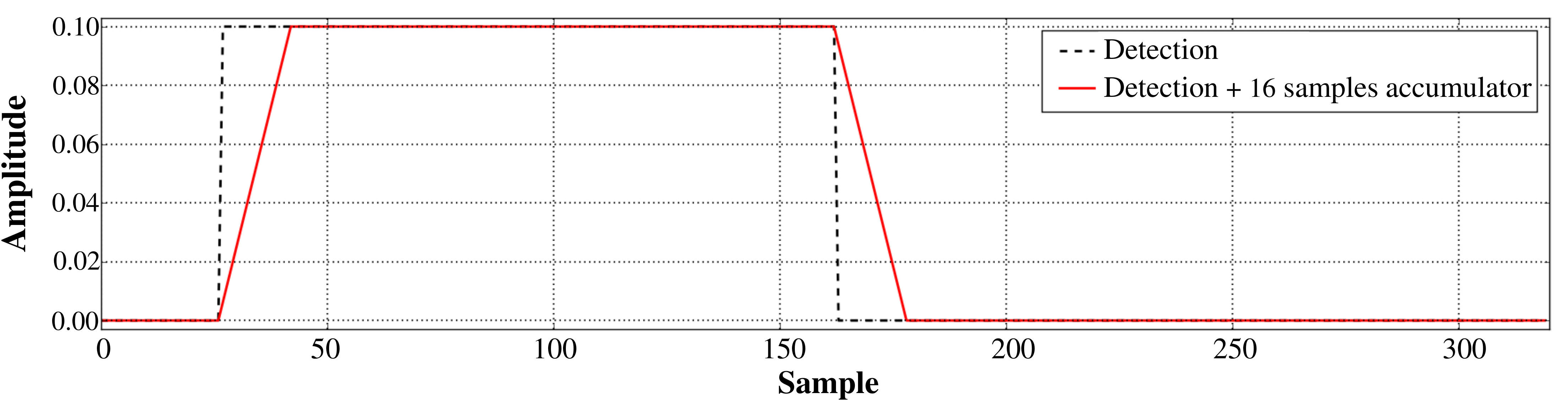}}\hfill
  \caption{(a) Autocorrelation waveform ${\left| {R\left[ n \right]} \right|^2}$ and signal power $P[n]^2$ (b) Frame detected ${{M[n]}}$> threshold.}
  \label{fig:frame_det_output} 
\end{figure}

\begin{figure}[tb]
  \centering
  \includegraphics[width=\columnwidth]{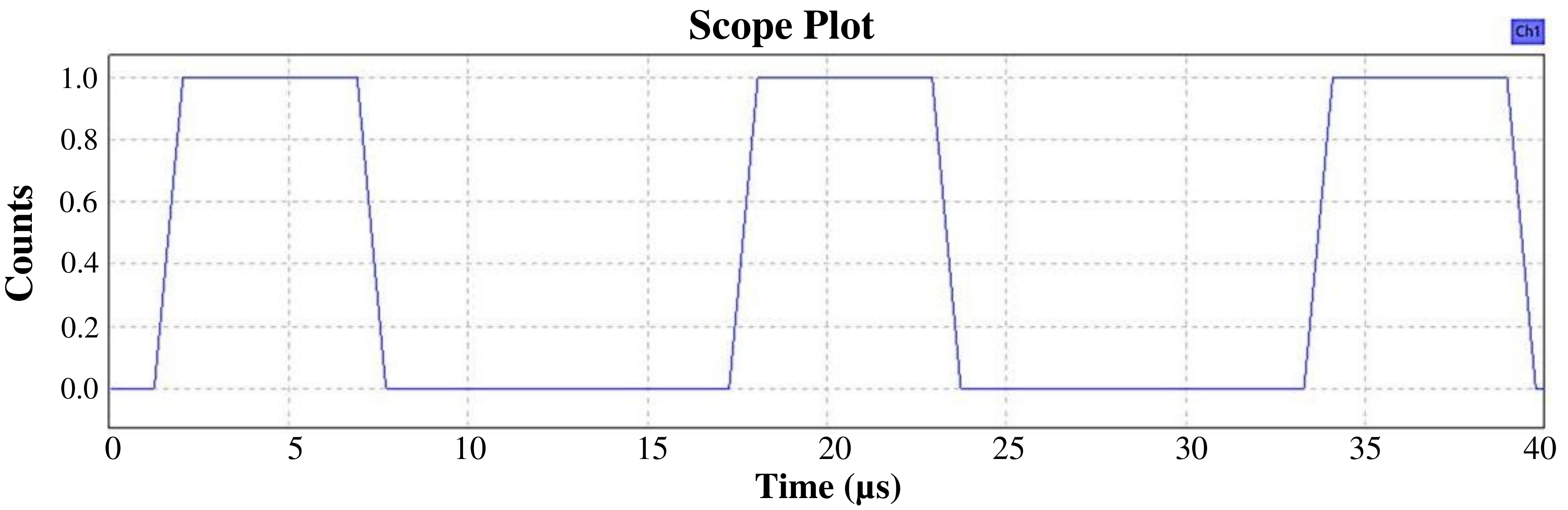} \vspace*{-1.0em}
  \caption{Frame detection output pulse on each signal received.}
  \label{fig:frame_det_output_pulse}
\vspace*{-1.0em}
\end{figure}

The frame detection scheme, based on \eqref{eq:rn_thr}, is shown in Fig. \ref{fig:frame_det_scheme}. The preamble length is 320 complex values, which is equivalent to a signal period of 16 $\mu s$. As shown in Fig. \ref{fig:frame_det_output_A}, the ratio between the absolute value of the correlation and the received power, indicates the presence of the frame at the receiver. Fig. \ref{fig:frame_det_output_B} shows the output of \eqref{eq:rn_thr}. Given that we sent several times the preamble through the channel, it is obtained a pulse train (see Fig. \ref{fig:frame_det_output_pulse}), which indicates the frames detected and the effectiveness of the implemented algorithm.

\section{Time synchronism}
\label{fig:time_synchronism}
Time synchronism is defined as the detection of the sample position where the frame or user information begins \cite{Barragan_2013a}. Like frame detection, this synchronism can be carried out by taking advantage of the preamble structure.

The implemented algorithm is based on the cross-correlation, and it is mathematically expressed as:
\begin{equation}
\Lambda [n] = \sum\limits_{m = 0}^{L - 1} {\left( {c_m^*{r_{n + m}}} \right)},
\label{eq:lambda_n}
\end{equation}
where $c_m^*$ is the complex conjugate of the preamble samples and the cross-correlator inputs are the samples of the received signal $r_n$. Such a cross-correlator needs $L$ complex multiplications to compute each output value, representing a high computational cost. However, this cost is compensated due to the results show that this detection is more precise than the autocorrelation algorithms.

The position $n_{xc\;\max }$ of the maximum value of $\left| {\Lambda [n]} \right|$, which provides the beginning of the user information, is defined as \cite{Xie_2010}:

\vspace*{-0.5em}
\begin{equation}
{n_{xc\;\max }} = \arg \mathop {\max } \left( {\left| {\Lambda [n]} \right|} \right).
\label{eq:n_xc_max}
\end{equation}

The time synchronism schemes that process the STS and LTS samples were implemented according to Fig. \ref{fig:time_synch_sch_STS} and Fig. \ref{fig:time_synch_sch_LTS}, respectively. In severe channel conditions, synchronization based on LTS is preferred because its output has a greater magnitude. Also, with STS, ten peaks have to be detected. Instead, with the LTS, only two peaks need to be identified, which implies a reduction in the algorithm's computational load.

The IEEE802.11a standard specifies that the autocorrelation peaks must be located at the sample number 160 for the STS and 320 for LTS. However, due to the channel noise and the propagation delay, a random displacement of the position of the symbol is expected. For this reason, equation \eqref{eq:n_xc_max} is performed within a window that is equal in size to the number of symbol samples.

\begin{figure}[t] 
  \centering
  \subfloat[\label{fig:time_synch_output_STS}]{
  \includegraphics[width=\columnwidth]{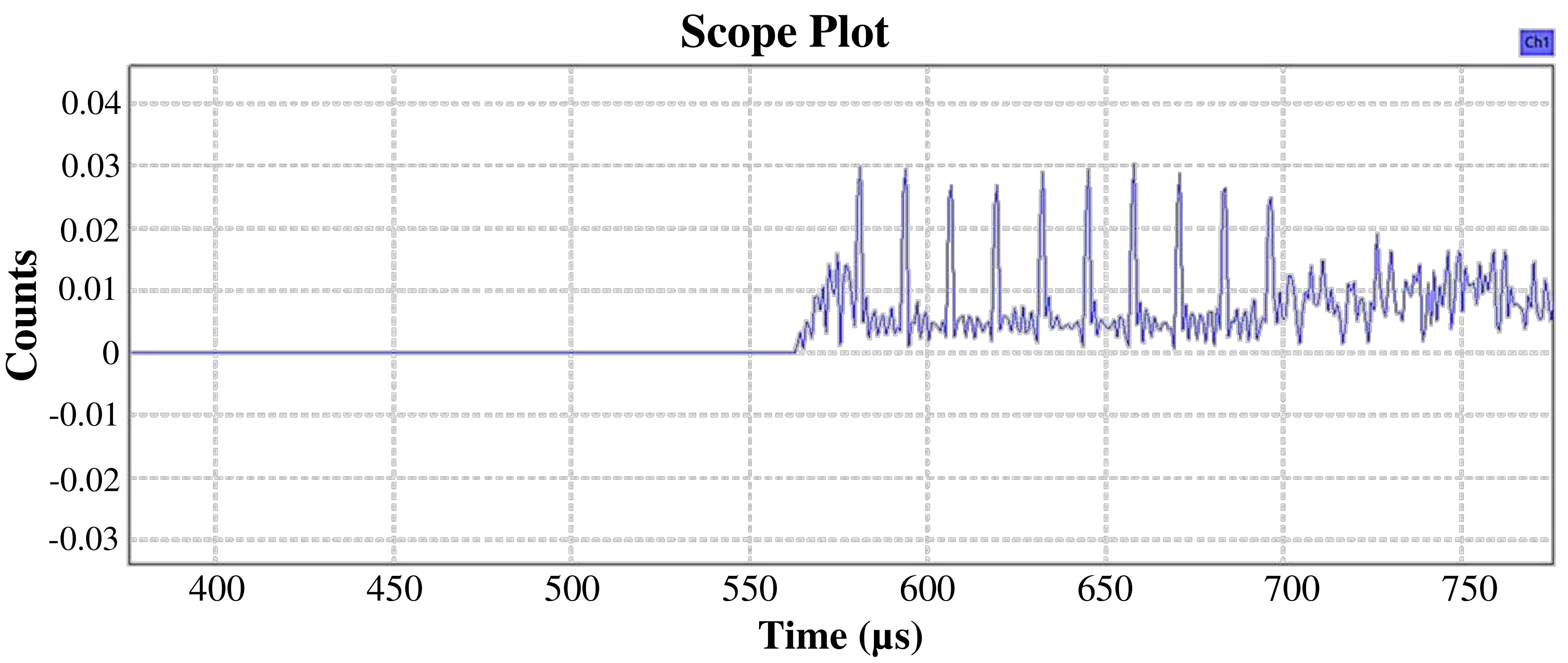}}\hfill
  \subfloat[\label{fig:time_synch_output_LTS}]{
  \includegraphics[width=\columnwidth]{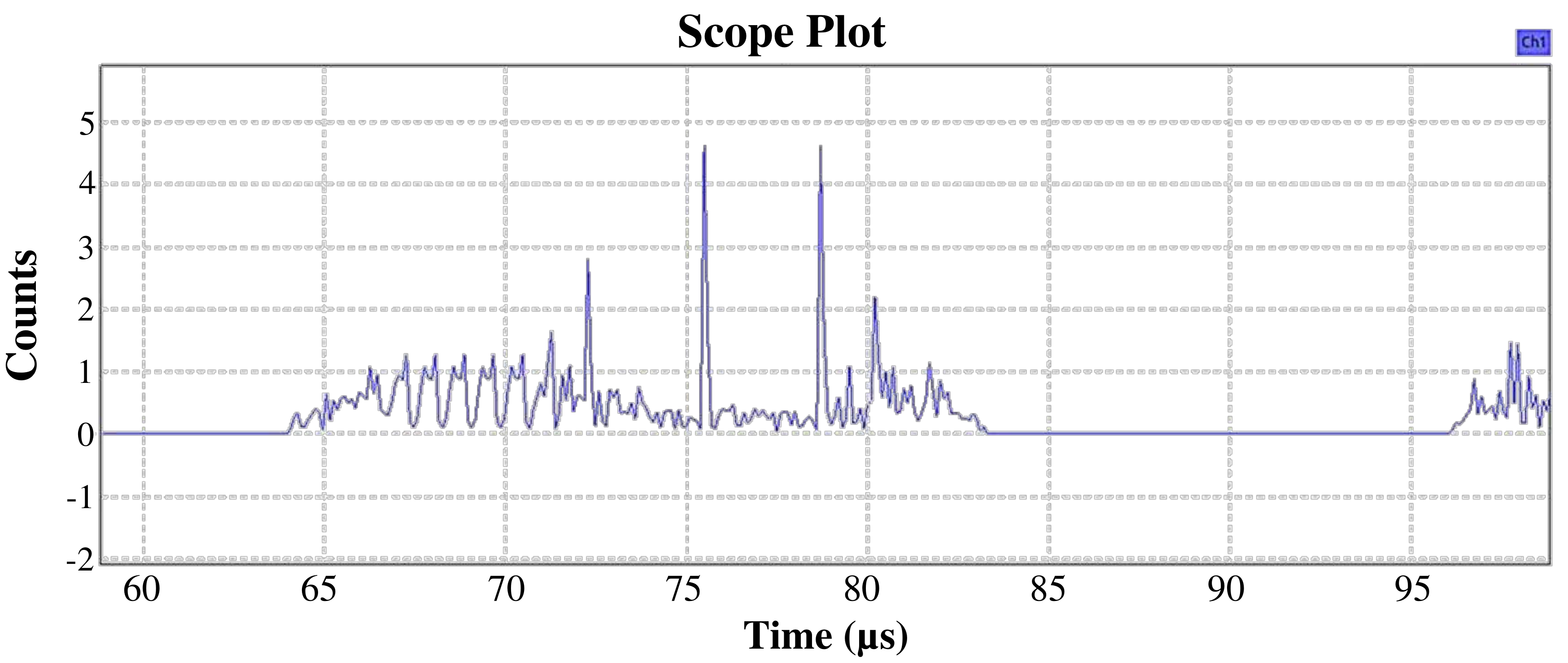}}\hfill
  \caption{Time synchronization scheme output with (a) STS and (b) LTS.}
  \label{fig:time_synch_output} 
\end{figure}

The results obtained after the implementation of the STS and LTS detectors are shown in Fig. \ref{fig:time_synch_output}. To obtain a correct correlation, zero inputs values at the beginning of each symbol (to complete the 320 signal samples) were added. Ten peaks corresponding to the STS symbols are shown in Fig. \ref{fig:time_synch_output_STS}; while in \ref{fig:time_synch_output_LTS} appears two peaks of the LTS symbols and the third one of smaller magnitude. The last peak is located at the ending position of the CP, and it is formed with the first 32 LTS samples.

\section{CFO detector and Frequency synchronism}
\label{fig:CFO_detector}
The main objectives of the frequency synchronism are to detect and compensate for the CFO ${\Delta _f}$ and the carrier phase error $\phi$. The ICI causes both impairments as a consequence of the Doppler effect, whose impact results in loss of orthogonality between subcarriers \cite{Chiueh_2007}. 

The Doppler effect, caused by the relative movement between transmitter and receiver, provokes a signal variation in time and frequency domains \cite{Cho_2010}. Thus, the frequency shift between the transmitted and received signal is computed by:

\vspace*{0.2em}
\begin{equation}
{\Delta _f} = \frac{\phi}{{2\pi n{T_s}}},
\label{eq:nueve}
\end{equation}
where $T_s$ is the sample time and $\phi  = 2\pi n{\Delta _f}$. 

When passing through the channel, the signal is affected by the ${\Delta _f}$ frequency offset. Therefore, the information reaches the receiver with a different carrier frequency (i.e., ${\Delta _f} = {f_{tx}} - {f_{rx}}$). The diagram used in this stage is shown in Fig. \ref{fig:CFO_scheme}. For simplicity, the effect of additive white Gaussian noise was not taken into account. 

To implement the CFO detector, the received signal $r_n$ from \eqref{eq:r_n} is replaced by the product between the discrete training sequence and a complex exponential with phase equal to ${\Delta _f}$. Hence, we obtain a simplified autocorrelation $R[n]$ defined as:

\begin{equation}
R[n] = \exp ( - j2\pi {\Delta _f}L{T_s})\sum\limits_{m = 0}^{L - 1} {({s_{n + m}}s_{n + m + L}^*)},
\label{eq:r_n_simp}
\end{equation}
where $s_n$ is a short symbol and $s_n^{*}$ is the complex conjugate.

The phase of the autocorrelation $R[n]$ is proportional to the phase offset $\phi$ introduced by the channel, i.e. $\phi \approx \arg (R[n])$. It allows finding the value of the CFO ${\Delta _f}$ through: 

\begin{equation}
{\Delta _f} = \frac{\arg (R[n])}{2\pi L T_s},
\label{eq:delta_f2}
\end{equation}
where $L = 16$.
Once the value of ${\Delta _f}$  is known, to compensate the frequency offset, the received signal is multiplied by a complex exponential with phase equal to ${-\Delta _f}$ as

\begin{equation}
{s_n} = {r_n}\exp ( - j2\pi {\Delta _f}n{T_s}).
\label{eq:trece}
\end{equation}

If ${\Delta _f}$ is similar to the frequency offset, the phase difference is compensated and the frame can be considered synchronized. 

\begin{figure}[tb] 
  \centering
  \subfloat[\label{fig:cfo_autocorr_output_30cm}]{
  \includegraphics[width=\columnwidth]{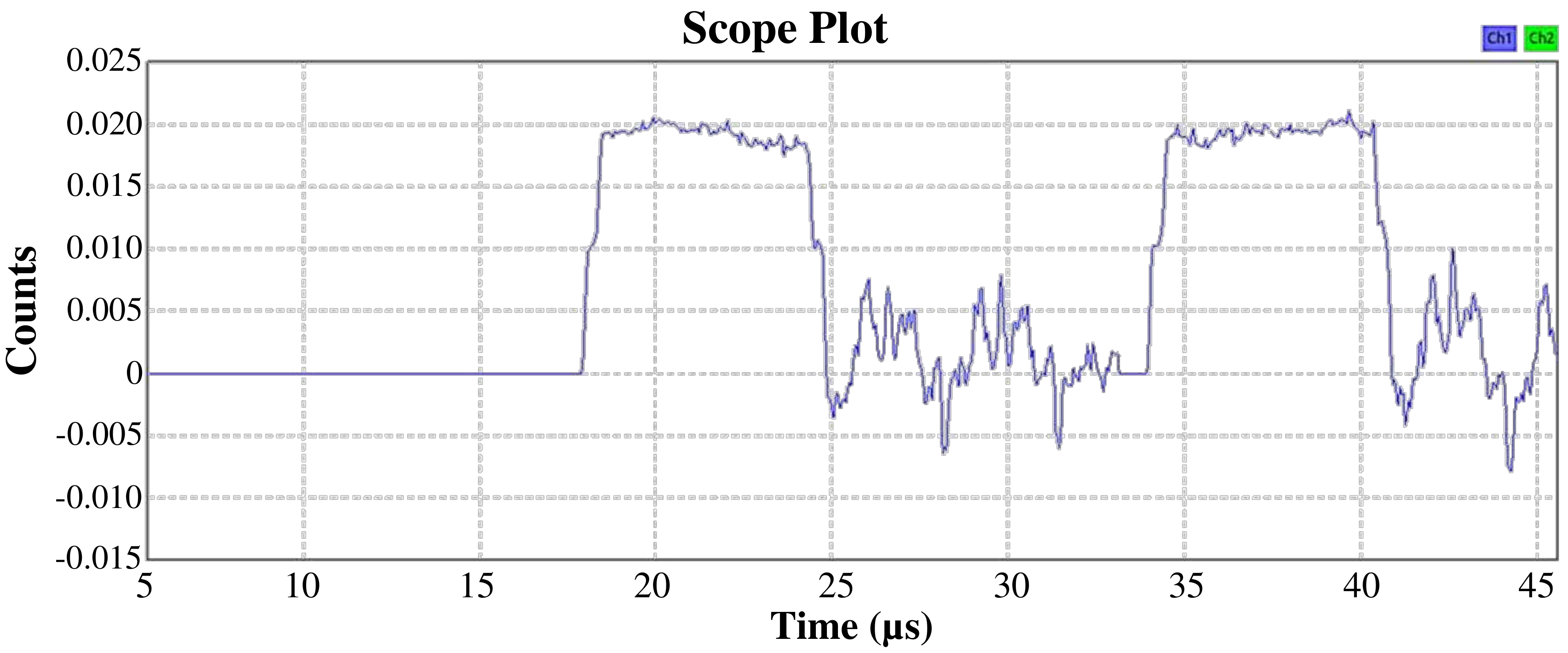}}\hfill
  \subfloat[\label{fig:cfo_autocorr_output_6m}]{
  \includegraphics[width=\columnwidth]{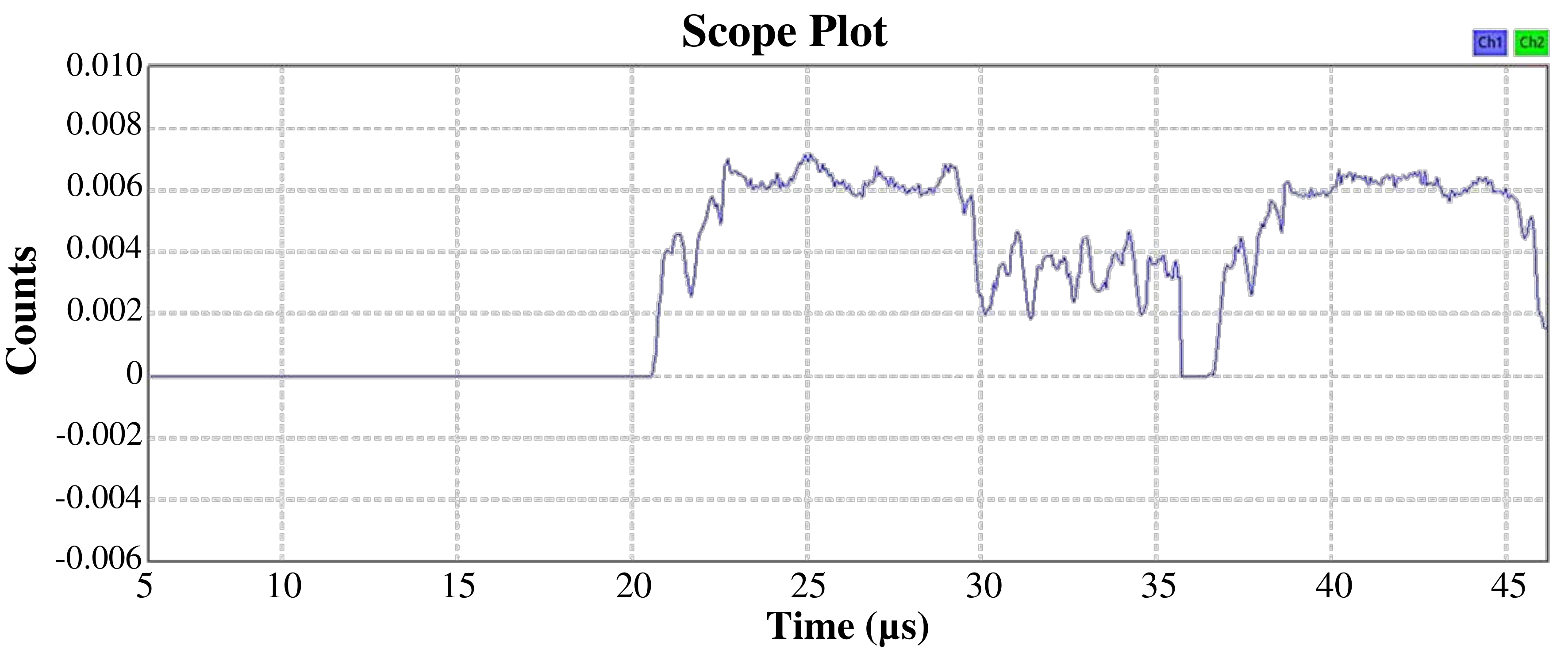}}\hfill
  \caption{Autocorrelation output used for the CFO detection at: (a) 0.3 m and (b) 6 m of distance between the USRPs.}
  \label{fig:cfo_autocorr_output} 
\end{figure}

Fig. \ref{fig:cfo_autocorr_output} shows the autocorrelation output used for the CFO detection. These waveforms change depending on the separation between the TX and the RX. Thus, in Fig. \ref{fig:cfo_autocorr_output_30cm} and Fig. \ref{fig:cfo_autocorr_output_6m} the distance was 0.3 m and 6 m, respectively.

\section{Results and discussion}
\label{fig:results_and_discussion}
Considering the frame detection scheme of Fig. \ref{fig:frame_det_scheme}, the transmitted continuous signal containing the preamble sequence was successfully detected in each transmission. The resulting waveform is an active signal (plateau-shaped) when at least two successive short symbols are correlated (32 first samples).

Each OFDM preamble was transmitted under slightly different channel conditions. Thus, it will result in different detected values for the time and frequency offset. The variance ${\sigma ^2}$ of the detected values $x_i$ is the metric used to assess the performance of the time and frequency synchronization algorithms. An algorithm whose variance is the smallest is more efficient. This variance is computed by:
\begin{equation}
{\sigma ^2} = \frac{{{\sum\limits_{i = 1}^N {{{({x_i} - \bar x)}^2}}}}}{N},
\label{eq:varianza}
\end{equation}
where $\bar x$ is the sample mean and $N$ is the number of trials.

\begin{figure}[t] 
  \centering
  \subfloat[\label{fig:histogram_time_synch_STS}]{
  \includegraphics[width=0.48\columnwidth]{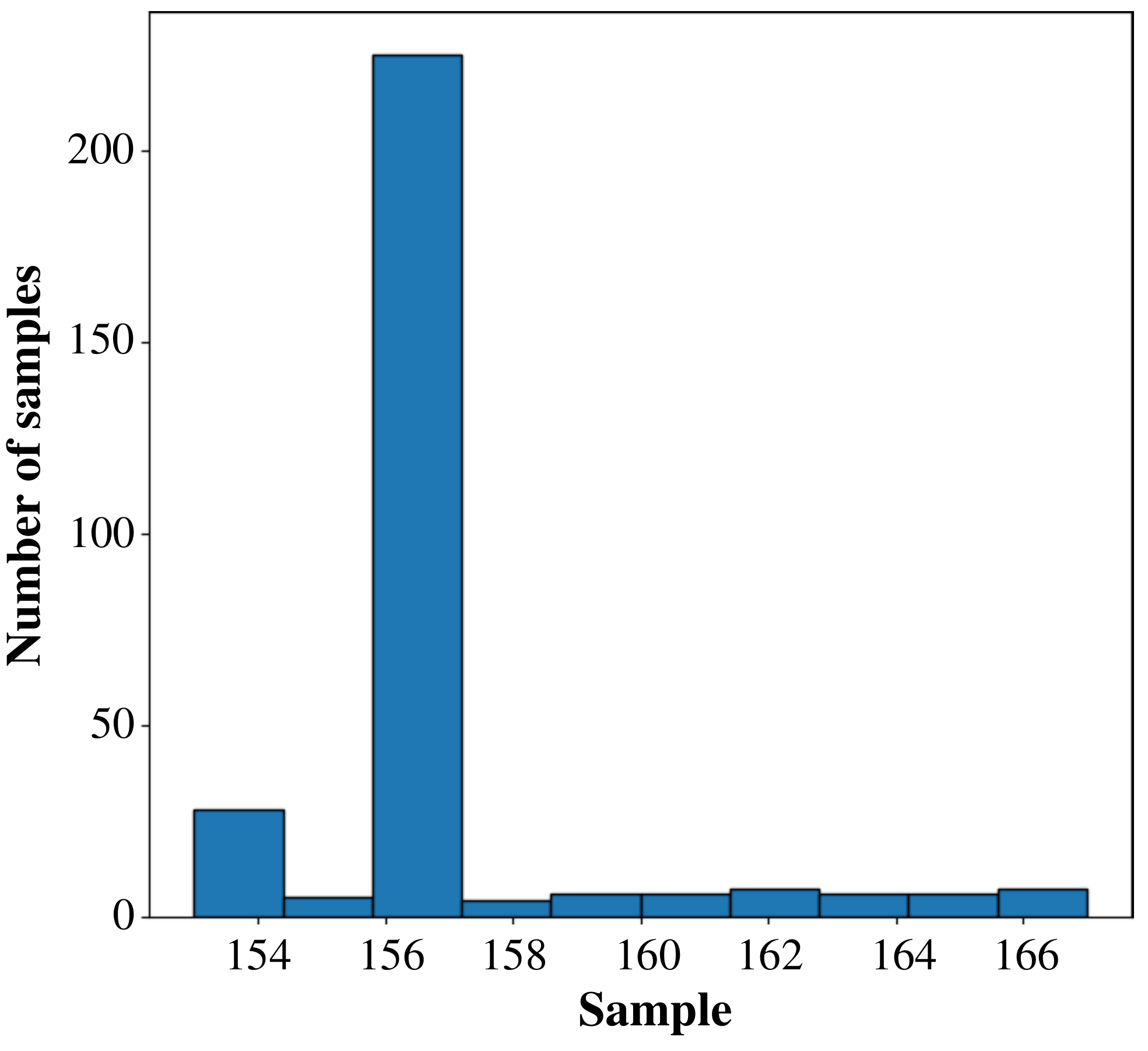}}\hfill
  \subfloat[\label{fig:histogram_time_synch_LTS}]{
  \includegraphics[width=0.48\columnwidth]{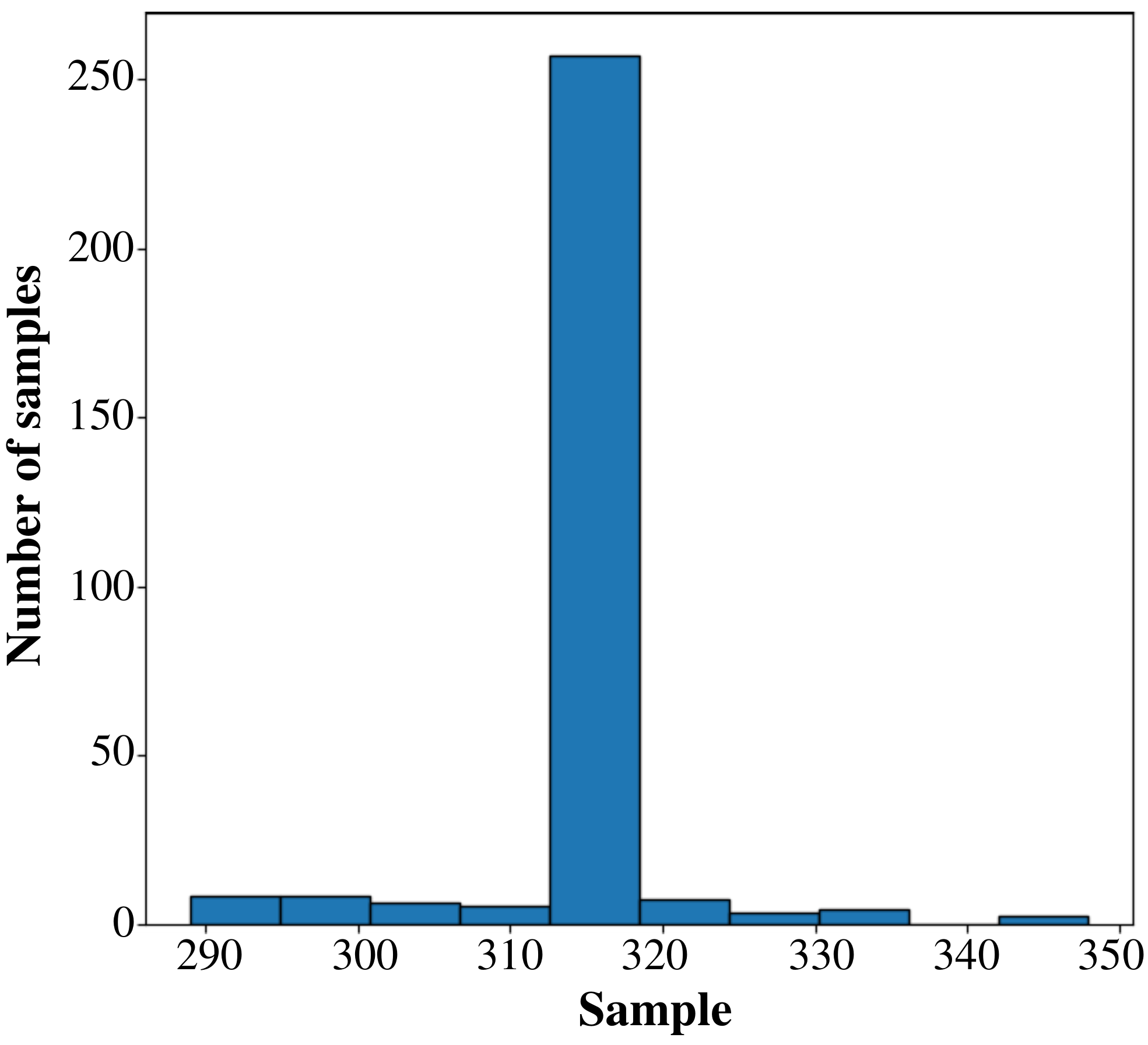}}
  \caption{Histogram of the synchronizer output using (a) STS symbols and (b) LTS symbols.}
  \label{fig:histogram_time_synch} 
\end{figure}

The obtained histograms for the STS and LTS synchronizers are shown in Fig. \ref{fig:histogram_time_synch_STS} and Fig. \ref{fig:histogram_time_synch_LTS}, respectively, which indicates an optimal performance of the implemented schemes, since they have a prominent value close to sample numbers 160 (STS) and 320 (LTS), as it is expected. The variance of the detected positions in this algorithm was computed and its result is shown in Table \ref{tab:table_2}.

According to Table \ref{tab:table_2}, the cross-correlation scheme using STS obtains a lower variance, making this scheme the most accurate algorithm. However, it is worth noting that in Fig. \ref{fig:time_synch_output_LTS}, the peak amplitude generated with the LTS cross-correlation is greater than the magnitude of the peaks with STS. In this way, the LTS cross-correlation output would be useful in severe channel conditions.

\begin{figure}[b] 
  \centering
  \subfloat[\label{fig:CFO_histogram_30cm}]{
  \includegraphics[width=0.48\columnwidth]{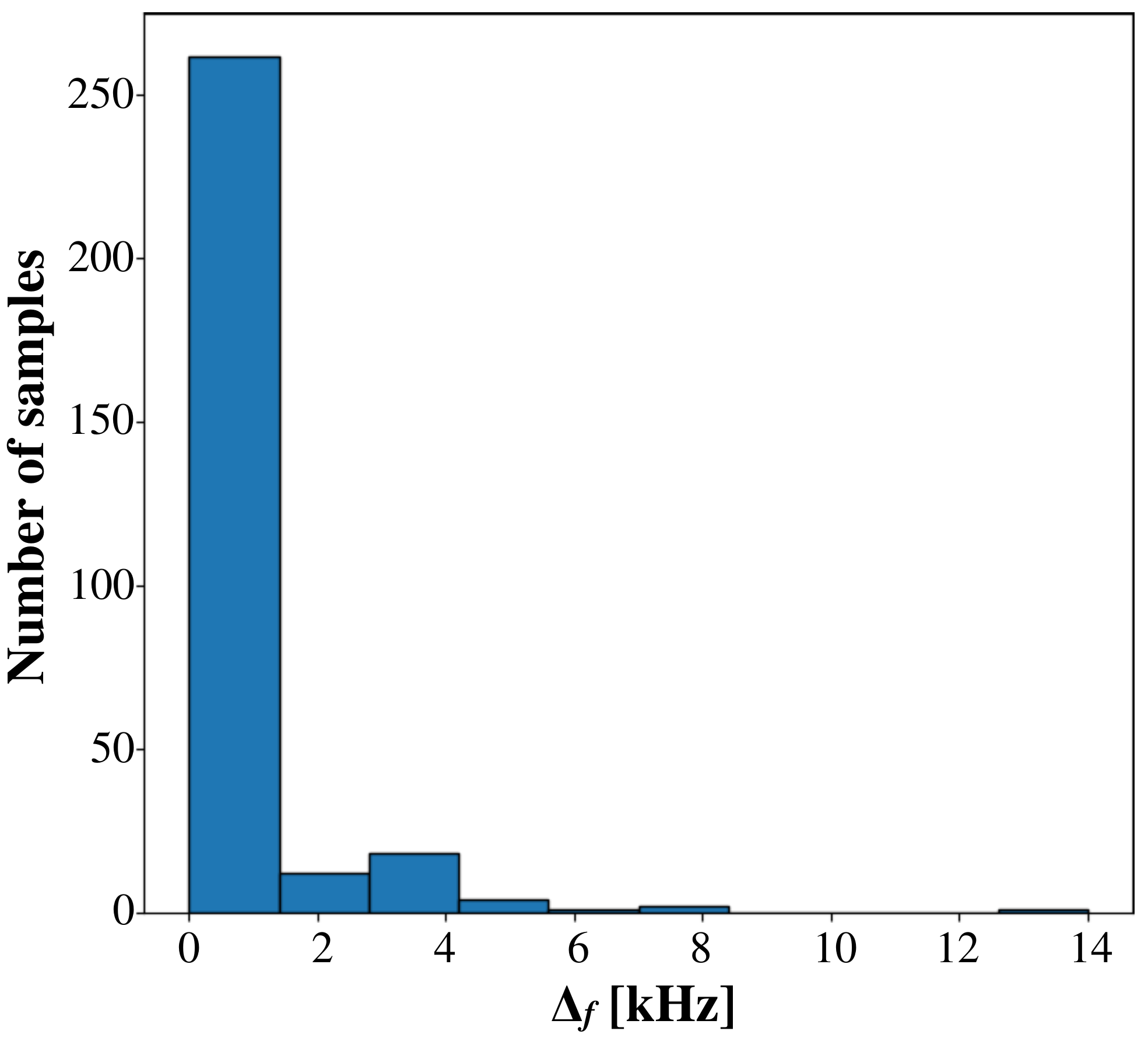}}\hfill
  \subfloat[\label{fig:CFO_histogram_6m}]{
  \includegraphics[width=0.48\columnwidth]{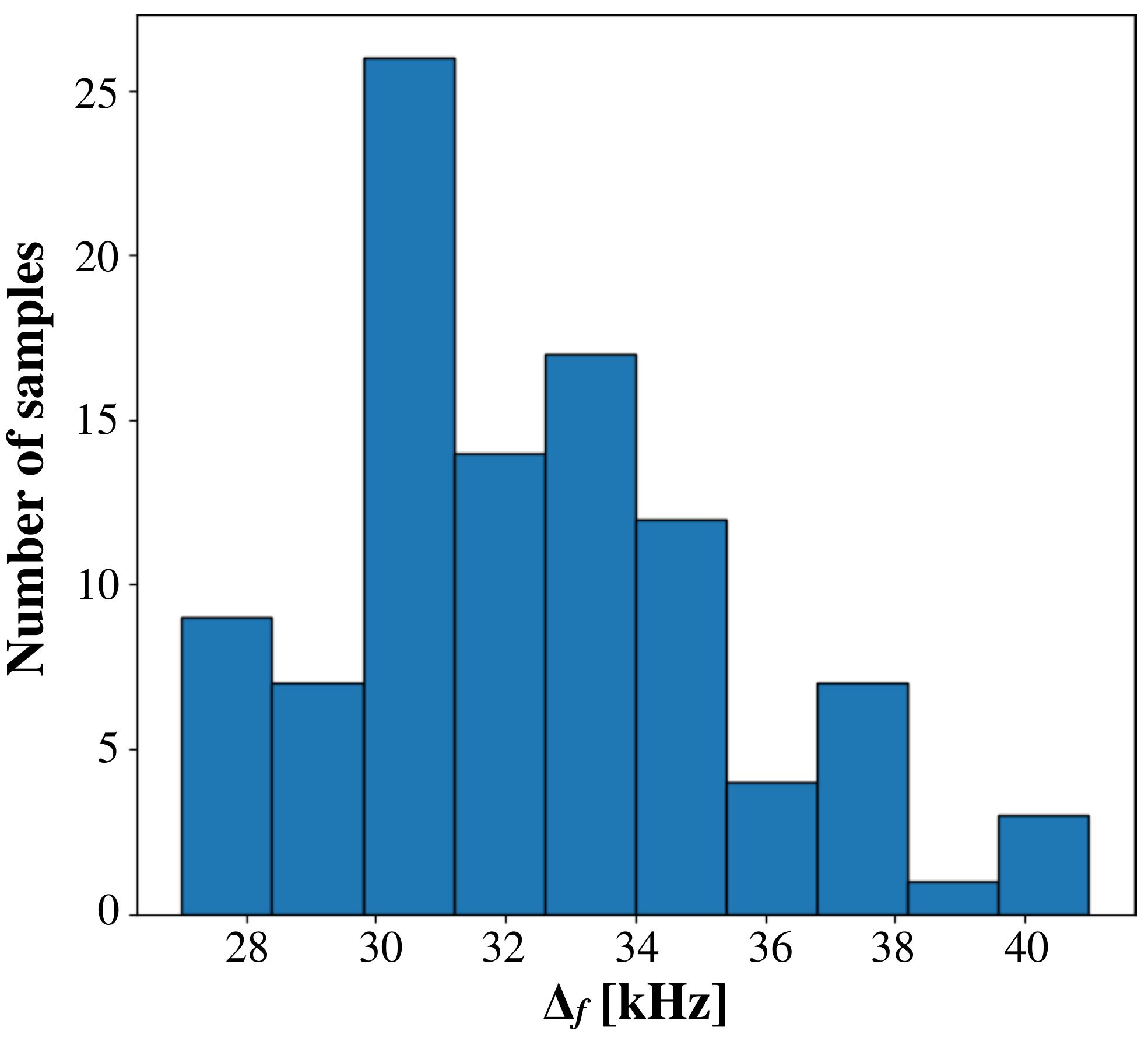}}
  \caption{Histogram of the CFO values detected at: (a) 0.3 m and (b) 6 m of distance between the USRPs.}
  \label{fig:CFO_histogram} 
\end{figure}

We performed two tests to evaluate the performance of the implemented CFO detector. In the first test, the USRPs were 0.30 m apart while in the second test, the USRPs were separated 6 m. In Fig. \ref{fig:CFO_histogram} the histograms of the CFO values detected with these two distances between the transmitter and the receiver are shown. We computed the variance of the detected frequency offset resulting in the numerical outcomes shown in Table \ref{tab:table_3}. As expected, the implemented algorithm detects a higher CFO at a distance of 6 m, with a prominent value of 30 kHz, which is then compensated using the scheme shown in Fig. \ref{fig:CFO_scheme}.
\begin{table}[t]
\renewcommand{\arraystretch}{1.3}
  \caption{VARIANCE OF TIME SYNCHRONIZATION ALGORITHMS}
  \centering
    \begin{tabular}{ccc} \hline 
    \textbf{Algorithms} & {\textbf{Trials (N)}} & $\mathbf{{\sigma ^2}}$ \bigstrut\\ \hline
    Cross-correlation  STS & 300   & 6.589  \bigstrut\\    \hline
    Cross-correlation  LTS & 300   & 45.040 \bigstrut\\    \hline
    \end{tabular}
  \label{tab:table_2}
  \vspace*{-1.0em}
\end{table}
\begin{table}[t]
\renewcommand{\arraystretch}{1.3}
  \centering
  \caption{VARIANCE OF THE FREQUENCY SYNCHRONIZATION ALGORITHM}
    \begin{tabular}{cccc}\hline
    \textbf{Algorithm} & {{\textbf{Distance TX-RX (m)}}} & {{\textbf{Trials (N)}}} &  $\mathbf{{\sigma ^2}}$ \bigstrut\\    \hline
    CFO detector & 0.3   & 300   & 2.200  \bigstrut\\    \hline
    CFO detector & 6     & 300   & 9.161  \bigstrut\\    \hline
    \end{tabular}
  \label{tab:table_3}
\end{table}

\section{Conclusions}
\label{section:sec_conclu}
In this paper, a preamble-based synchronizer implemented in USRP for OFDM is reported. The OFDM synchronism is divided into three stages: the packet or frame detection, time synchronism, and frequency offset detection and correction. In frame detection, the autocorrelation amplitude's modulus was compared against the average power of the signal multiplied by a threshold. When the autocorrelated signal's modulus was higher than half of the average power in at least 32 samples (equivalent to two STS symbols), the result was a step-type (plateau-shaped) active signal indicating the successful detection in the receiver of a transmitted frame.

In time synchronism, the wireless channel noise caused a variation in the positions of the peaks of STS and LTS correlation, as expected. Based on the results, cross-correlation with STS yields a more accurate synchronism. Besides, it is recommended to use in severe channel environments the cross-correlation scheme with the LTS.

According to the practical experiments, the CFO detected with the implemented scheme is proportional to the separation between the transmitter and receiver. At 0.3 m distance between the USRPs, the frequency offset was close to 0 kHz. As the distance between the USRPs raised, the CFO values were increased with each performed test. With a separation of 6 m in an indoor environment, an average of 32 kHz of frequency offset was detected.

The implemented synchronization algorithms offer promising results over the AWGN channel for an OFDM system, where it was shown to be suitable for preamble-based detectors. Finally, extending the proposed implementation to other 802.11 amends would be of interest for future research.

\balance
\bibliographystyle{IEEEtran}
\bibliography{IEEEabrv,ConfAbrv,GP4}
\end{document}